\documentclass[aps,amssymb,pra,showpacs,twocolumn]{revtex4}

\usepackage{epsf}
\usepackage{amsmath}
\usepackage{graphicx}
   
\begin{document}                                                       
                                            
\title{Cyclotron enhancement of tunneling}

\author{M. V. Medvedeva,$^{1}$ I. A. Larkin,$^{2,3}$ S. Ujevic,$^{2}$ L. N. Shchur,$^{1}$ and  B. I. Ivlev,$^{4,5}$} 

\affiliation
{$^{1}$Landau Institute for Theoretical Physics, Moscow, Russia,\\
$^{2}$International Center of Condensed Matter Physics, Bras\'ilia, Brazil\\
$^{3}$Institute of Microelectronics Technology, Chernogolovka, Russia\\
$^{4}$Universidad Aut\'onoma de San Luis Potos\'{\i}, San Luis Potosi, Mexico,\\
$^{5}$Department of Physics and Astronomy and NanoCenter, University of South Carolina, Columbia, South Carolina, USA}


\begin{abstract} 
A  state of an electron in a quantum wire or a thin film becomes metastable, when a static electric field is applied 
perpendicular to the wire direction or the film surface. The state decays via tunneling through the created potential 
barrier. An additionally applied magnetic field, perpendicular to the electric field, can increase the tunneling decay rate
for many orders of magnitude. This happens, when the state in the wire or the film has a velocity perpendicular to the 
magnetic field. According to the cyclotron effect, the velocity rotates under the barrier and becomes more aligned with the
direction of tunneling. This mechanism can be called cyclotron enhancement of tunneling.
\end{abstract} \vskip 1.0cm
   
 
\maketitle

\section{INTRODUCTION}
\label{sec:intr}
Tunneling in a magnetic field is a matter of investigation for many years. The magnetic field can influence tunneling
across a potential barrier in two different ways. 

First, the magnetic field can modify an underbarrier motion related to a classically forbidden region. Studies of this 
phenomenon are presented in the literature. In Refs.~\cite{SHKL1,SHKL2} it was pointed out that an underbarrier fall of the
wave function can be less rapid in an inhomogeneous sample. See also Refs.~\cite{THOUL,BLATT,GOROKH,KHAET,RAIKH}. In 
Ref.~\cite{HALP} a transmission coefficient through a quadratic barrier was found. A decay of a metastable state was 
considered in Ref.~\cite{DYK}. The certain peculiarities of an underbarrier wave function were discussed in 
Refs.~\cite{IVLEV1,IVLEV2}. 

Second, the magnetic field can influence a state of an electron at a classically allowed region after an exits from under
the barrier. A typical example is the Wigner resonance when the electron tunnels into a potential well with a level
aligned to its energy \cite{LANDAU}. See experimental measurements, for instance, in Refs.~\cite{DEMM,BET,VDOV1}. Another 
example relates to a specific density of states in the classical region after the tunneling barrier. A state of an 
electron, influenced by the magnetic field, may fit better that density of states and this results in increase of tunneling
rate \cite{VDOV2}. 

The goal of the paper is to study tunneling decay rate of a metastable state in a magnetic field (the electron after 
tunneling goes to infinity). The question to be answered: Can a magnetic field increase the decay rate? It is clear that the
above effect of density of states at the region after the barrier can, in principle, increase the rate. But this effect,
related to a prefactor, cannot be very large. According to Ref.~\cite{VDOV2}, there is approximately $50\%$ enhamcement of
the effect. 

It would be much more amazing to increase the main exponential part of the decay rate by the magnetic field. The 
references \cite{SHKL1,SHKL2,THOUL,BLATT,GOROKH,KHAET,RAIKH,HALP,DYK} say that it is impossible. Indeed, when an electron 
enters under the barrier its velocity deviates, due to the cyclotron effect, from a tunneling path with no magnetic field. 
This leads to a reduction of the tunneling probability by the magnetic field. The reduction can be also explained in terms
of increasing of the total barrier. The additional barrier is proportional to a squared velocity of the electron in the 
magnetic field \cite{SHKL1,SHKL2}. 

But there is a situation when the electron tunnels from a quantum wire or another object extended in the direction 
perpendicular to tunneling. In this case a state prior to tunneling can have a finite velocity perpendicular to the 
tunneling direction. According to the cyclotron effect, this velocity rotates under the barrier and becomes more aligned 
with the tunneling direction. This leads to enhancement of the tunneling rate by the magnetic field (cyclotron 
enhancement). 

Formally, cyclotron enhancement of tunneling results from a reduction of the main tunneling exponent which reminds one of 
Wentzel, Kramers, and Brillouin (WKB). The exponent can be reduced in a few times. Suppose that at zero magnetic field the 
tunneling rate is proportional to $10^{-24}$. The magnetic field can turn it into, say, $10^{-7}$. 

We consider in the paper tunneling from a straight quantum wire, directed in the $y$ axis, embedded into a two-dimensional
electron system in the $\{x,y\}$ plane. The potential barrier is created by the electric field ${\cal E}_{0}$ directed 
along the $x$ axis (the direction of tunneling). The magnetic field $H$ is aligned along $z$. According to electrodynamics,
after tunneling a motion of the electron in perpendicular magnetic and electric fields is restricted by a finite interval 
in the $x$ direction \cite{LANDAU1}. To get the electron passed to the infinite $x$ one should put some potential wall(s) 
along the $x$ direction restricting the $y$ motion. Due to multiple reflections from the restricting wall in the magnetic 
field the electron goes to the infinite $x$. We model the walls by the potential proportional to $y^{4}$. 

The theory presented relates also to tunneling from a flat $\{y,z\}$ film with quantized electron motion in the $x$ 
direction. The electron tunnels into a three-dimensional reservoir. Restricting walls should be placed parallel to the 
$\{x,z\}$ plane. 

Without the restricting walls a solution can be obtained analytically on the bases of the modified WKB approach as shown in
Sec.~\ref{sec:wkb}. An approximation of classical complex trajectories is formulated in Sec.~\ref{sec:traj}. In 
Secs.~\ref{sec:masha} and \ref{sec:vania} two different methods of numerical calculations are applied to the problem with 
restricting walls. 
\section{FORMULATION OF THE PROBLEM}
\label{sec:form}
We consider an electron localized in the $\{x,y\}$ plane. The static magnetic field $H$ is directed along the $z$ axis. 
Suppose a motion of the electron in the $\{x,y\}$ plane to occur in the potential $U(x,y)$. Then the Schr\'{o}dinger 
equation, with the vector potential $\vec A=\{0,Hx,0\}$, has the form \cite{LANDAU}
\begin{equation} 
\label{1}
-\frac{\hbar^{2}}{2m}\frac{\partial^{2}\psi}{\partial x^{2}}
-\frac{\hbar^{2}}{2m}\left(\frac{\partial}{\partial y}+ix\frac{m\omega_{c}}{\hbar}\right)^{2}\psi+U(x,y)\psi=E\psi,
\end{equation}
where $\omega_{c}=|e|H/mc$ is the cyclotron frequency. The potential 
\begin{equation} 
\label{2}
U(x,y)=-\hbar\sqrt{\frac{2u_{0}}{m}}\delta(x)-{\cal E}_{0}x+u_{0}\frac{y^{4}}{a^{4}}
\end{equation}
describes the quantum wire placed in the $y$ direction (the first term), the constant electric field ${\cal E}_{0}$ (the 
second term), and the restricting walls in the $y$ direction are modeled by the third term. At ${\cal E}_{0}=0$ and 
$H=0$ the discrete energy level in the $\delta$ well ($-u_{0}$) is a ground state in the WKB approximation.
\subsection{Dimensionless units}
Let us introduce the dimensionless electric field $\varepsilon$ and the magnetic field $h$ by the equations
\begin{equation} 
\label{24}
\varepsilon=\frac{a{\cal E}_{0}}{u_{0}},\hspace{1cm}h=\frac{\omega_{c}}{{\cal E}_{0}}\sqrt{\frac{mu_{0}}{2}}.
\end{equation}
Below we measure $x$ and $y$ in the units of $u_{0}/{\cal E}_{0}$ and time in the units of
\begin{equation} 
\label{25}
\tau_{00}=\frac{\sqrt{2mu_{0}}}{{\cal E}_{0}}.
\end{equation}
The energy is $E=u_{0}\lambda$ where the dimensionless energy $\lambda$ is negative in our problem. We also introduce a 
large semiclassical parameter
\begin{equation} 
\label{26}
B=\frac{u_{0}\sqrt{2mu_{0}}}{\hbar {\cal E}_{0}}.
\end{equation}
At zero magnetic field $(h=0)$ the WKB probability of tunneling \cite{LANDAU}
\begin{equation} 
\label{26b}
w(h=0)\sim\exp\left(-\frac{4B}{3}\right)
\end{equation}
is small. In the new variables Schr\"{o}dinger equation (\ref{1}) has the form
\begin{eqnarray} 
\label{26a}
&&-\frac{1}{B^{2}}\frac{\partial^{2}\psi}{\partial x^{2}}
-\left(\frac{1}{B}\frac{\partial}{\partial y}+ihx\right)^{2}\psi\\
\nonumber
&&+\left[\frac{y^{4}}{\varepsilon^{4}}-x-\frac{2}{B}\delta(x)\right]\psi=\lambda\psi.
\end{eqnarray}
One can exclude the point $x=0$, if to impose the boundary condition 
\begin{equation} 
\label{2a}
\frac{\partial\psi}{\partial x}\bigg|_{x=+0}-\frac{\partial\psi}{\partial x}\bigg|_{x=-0}=-2B\psi(0,y).
\end{equation}
With the semiclassical accuracy, the lowest level at the $\delta$ well is $\lambda=-1$. At a finite electric field that 
level is metastable and we should find a decay rate due to tunneling in the magnetic field.
\subsection{A semiclassical approach}
\label{sec:semi}
When the potential barrier is hardly transparent (a large $B$) one can use the semiclassical approximation for the wave 
function \cite{LANDAU}
\begin{equation} 
\label{3}
\psi(x,y)\sim\exp\left[iB\sigma(x,y)\right],
\end{equation}
where $\sigma(x,y)$ is the classical action satisfying the equation of Hamilton-Jacobi at $x\neq 0$
\begin{equation} 
\label{4}
\left(\frac{\partial\sigma}{\partial x}\right)^{2}
+\left(\frac{\partial\sigma}{\partial y}+hx\right)^{2}-x+\frac{y^{4}}{\varepsilon^{4}}=\lambda.
\end{equation}
The function $\sigma(x,y)$ is continuous. As follows from Eq.~(\ref{4}), $\left(\partial\sigma/\partial x\right)^{2}$ is 
also continuous. According to that, one can consider the action at positive $x$ with the boundary condition
\begin{equation} 
\label{4a}
\frac{\partial\sigma}{\partial x}\bigg|_{x=0}=i,
\end{equation}
which follows from Eq.~(\ref{2a}). In the approximation of a large $B$ the condition (\ref{2a}) is reduced to
\begin{equation} 
\label{4b}
\frac{\partial\psi(x,y)}{\partial x}\bigg|_{x=0}=-B\psi(0,y),
\end{equation}
if to consider the problem at positive $x$ only.
\section{CYCLOTRON ENHANCEMENT OF TUNNELING}
\label{sec:wkb}
First, we consider tunneling from the quantum wire, when there are no restriction of the motion in the $y$ direction. In 
other words, we drop down the term $y^{4}$ in the potential $U(x,y)$ (\ref{2}). In this case the solution of the 
Schr\"{o}dinger equation (\ref{1})
\begin{equation} 
\label{36a}
\psi(x,y)=\exp\left(-iBky\right)\varphi(x)
\end{equation}
is determined by the effective Schr\"{o}dinger equation
\begin{equation} 
\label{36b}
-\frac{1}{B^{2}}\frac{d^{2}\varphi}{dx^{2}}+V(x)\varphi=\lambda\varphi
\end{equation}
with the effective potential energy 
\begin{figure}
\includegraphics[width=5.5cm]{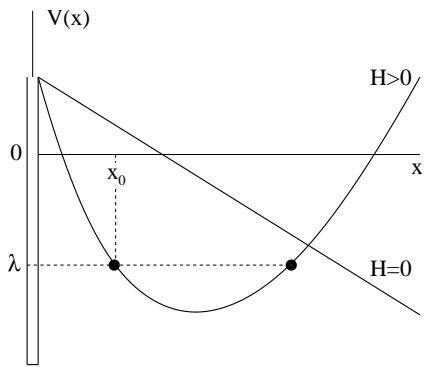}
\caption{\label{fig1}The effective potential energy (\ref{36c}) is plotted for a positive $k$ in Eq.~(\ref{36d}). An 
electron tunnels between $x=0$ and $x_{0}$. The two dots mark terminal points between which a classical motion occurs. The 
potential barrier with no magnetic field is shown by the straight line.}
\end{figure}
\begin{equation} 
\label{36c}
V(x)=\left(k-hx\right)^{2}-x-\frac{2}{B}\delta(x).
\end{equation}
The first term in Eq.~(\ref{36c}) is proportional to squared velocity of the electron. Solutions with positive and negative 
wave vector $k$ are possible according to different directions of motion in the wire. In the WKB approximation for the 
state, localized in the $\delta$ well, 
\begin{equation} 
\label{36d}
k=\pm\sqrt{1+\lambda}.
\end{equation}
In the physical units, the velocity along the quantum wire is
\begin{equation} 
\label{36e}
v_{y}=\mp\sqrt{1+\lambda}\sqrt{\frac{2u_{0}}{m}}.
\end{equation}
\begin{figure}
\includegraphics[width=7cm]{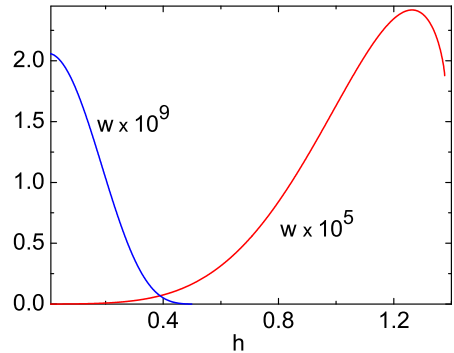}
\caption{\label{fig2}(Color online) The decay rate (\ref{41}) of the metastable state of the quantum wire versus  magnetic 
field $h$ at $B=15$. Left curve: The state in the wire has zero velocity parallel to the wire, $\lambda=-1$. Right
curve: The same when the velocity is finite, $\lambda=-0.594$.} 
\end{figure}
Below we consider a positive $k$ related to the motion from $+\infty$ to $-\infty$ in $y$. For this case the potential 
energy $V(x)$ is shown schematically in Fig.~\ref{fig1} and the WKB wave function under the barrier, $0<x<x_{0}$, has the 
form
\begin{eqnarray} 
\label{37}
&&\psi(x,y)\sim\exp\left(-iyB\sqrt{1+\lambda}\right)\\
\nonumber
&&\exp\left[-B\int^{x}_{0}dx_{1}\sqrt{\left(\sqrt{1+\lambda}
-hx_{1}\right)^{2}-x_{1}-\lambda}\hspace{0.05cm}\right].
\end{eqnarray}
The classical terminal point, determined by the condition $V(x)=\lambda$, is 
\begin{equation} 
\label{38}
x_{0}=\frac{1}{h}\left(\sqrt{1+p^{2}}-p\right),
\end{equation}
where 
\begin{equation} 
\label{39}
p=\sqrt{\left(\frac{1}{2h}+\sqrt{1+\lambda}\right)^{2}-1}.
\end{equation}
The terminal point $x_{0}$ plays a role of the exit point from under the barrier and is marked in Fig.~\ref{fig1}. The 
classically allowed region, $V(x)<\lambda$, is between two terminal points displayed in Fig.~\ref{fig1}. Our consideration 
has a meaning, when the classically allowed region is of a finite width. This occurs under the condition
\begin{equation} 
\label{40}
h<\frac{1}{2(1-\sqrt{1+\lambda})},
\end{equation}
when the two terminal points in Fig.~\ref{fig1} are separated. 

For the $y$-independent potential considered the classically allowed motion occurs between the two terminal points in 
Fig.~\ref{fig1}. It means that after an exit from under the barrier the electron motion will remain restricted in the $x$
direction. A penetration through the barrier, $0<x<x_{0}$, in Fig.~\ref{fig1} can be enhanced, when the level $\lambda$ in
the $\delta$ well coincides with one of the discrete levels in the well at $x>x_{0}$. The Wigner resonant tunneling occurs
in this case \cite{LANDAU}. 

A scenario becomes different, if we return to the full potential energy (\ref{2}). After coming from under the barrier the
electron participates in cyclotron motion and the Lorentz drift similar to a classical particle. The Lorentz drift in the 
potential $(u_{0}y^{4}/a^{4}-{\cal E}_{0}x)$ brings the electron to the infinite $x$. This happens, according to  
electrodynamics, as a result of multiple reflections from the ``wall'' $u_{0}y^{4}/a^{4}$ \cite{LANDAU1}. In that case 
Wigner tunneling does not occur since there are no discrete levels to the right of the barrier. This situation corresponds
to a decay of the metastable state in the $\delta$ well since after tunneling the electron goes to the infinite $x$.  

Whereas the potential $y^{4}$ strongly disturbs the motion at the classical region, the underbarrier motion along the
direction $y=0$ is hardly violated since reflections from that potential under the barrier are less important. It is shown 
in Sec.~\ref{sec:masha}. For this reason, for an underbarrier motion in the full potential (\ref{2}) one can use 
the approach, when the potential $y^{4}$ is dropped down. Accordingly, in the semiclassical approximation the tunneling 
rate is given by the WKB formula
\begin{equation} 
\label{41}
w\sim\exp\left[-\frac{B}{h}\left(\sqrt{1+p^{2}}-p^{2}\ln\frac{1+\sqrt{1+p^{2}}}{p}\right)\right],
\end{equation}
which follows from Eq.~(\ref{37}) as $|\psi(x_{0},y)|^{2}$.

For zero magnetic field, $h=0$, the tunneling rate (\ref{41}) does not depend on $\lambda$ and coincides with the 
conventional WKB expression (\ref{26b}). For $B=15$ and $\lambda=-1$ the tunneling rate in Fig.~\ref{fig2} drops down with 
the magnetic field, as usually, since the metastable state in the quantum wire has zero momentum (\ref{36d}) perpendicular 
to tunneling. For $B=15$ and $\lambda>-1$ such a momentum is finite resulting in cyclotron enhancement of tunneling. In
Fig.~\ref{fig2} we plot the tunneling rate for $\lambda=-0.594$. This value is chosen to have a link to numerical data
discussed below.
\section{CLASSICAL TRAJECTORIES}
\label{sec:traj}
For the potential (\ref{2}) analytical solutions of the Schr\"{o}dinger equation (\ref{26a}) and the Hamilton-Jacobi 
equation (\ref{4}) do not exist. Nevertheless, to calculate a tunneling rate with the exponential accuracy it is not 
necessary to solve the Hamilton-Jacobi equation (\ref{4}) in the full $\{x,y\}$ plane. It is sufficient to track 
$\sigma(x,y)$ along a classical trajectory $\{x(\tau),y(\tau)\}$, where $\tau=-it$ is imaginary time since an underbarrier 
classical motion is impossible. 

The method of classical trajectories in imaginary time is well developed for static potential barrier with no magnetic 
field \cite{COLEMAN1,COLEMAN2,SCHMID1,SCHMID2}. In this case the main contribution to a tunneling rate comes from the 
extreme path linking two classically allowed regions. The real underbarrier path $\{x(\tau),y(\tau)\}$ can be parametrized 
as a classical trajectory in imaginary time. 

A situation with a magnetic field is substantially different. For the potential (\ref{2}) in a magnetic field the 
coordinate $x(\tau)$ remains real in the underbarrier motion but $y(\tau)=-i\eta(\tau)$ becomes imaginary. This means that 
the trajectory does not track the entire underbarrier path as for zero magnetic field. It just provides a ``bypass''  
through the plane of complex $y$. So the method of classical trajectories for tunneling in a magnetic field is nontrivial. 

One should note that the same situation takes place for tunneling across nonstationary barriers where imaginary time stands 
instead of imaginary $y$. Validity of the nontrivial trajectory method for the nonstationary barrier was numerically 
proved in Ref.~\cite{IVLEV8}. The numerical results of the present paper support the trajectory method also for a static 
magnetic field. 

We consider tunneling from the $\delta$ well ($\tau=\tau_{0}$) to the classically allowed region ($\tau=0$). The ratio of 
the densities
\begin{equation} 
\label{5}
w=\frac{|\psi(x(0),0)|^{2}}{|\psi(0,0)|^{2}}
\end{equation}
can be identified with a tunneling rate. With the semiclassical accuracy, $w\sim w_{T}$, where
\begin{equation} 
\label{14}
w_{T}=\exp\left(-2B{\rm Im}\left[\sigma(x(0),0)-\sigma(0,0)\right]\right)
\end{equation}
is the tunneling exponent. The trajectory method allows to calculate the part 
\begin{equation} 
\label{15}
{A}_{0}=2B{\rm Im}\left[\sigma(x(0),0)-\sigma(0,-i\eta(i\tau_{0}))\right]
\end{equation}
of the total action (\ref{14}) only. This part is expressed through the classical trajectory
\begin{eqnarray} 
\label{27}
&&A_{0}=2B\int^{\tau_{0}}_{0}d\tau\bigg[\frac{1}{4}\left(\frac{\partial x}{\partial\tau}\right)^{2}-
\frac{1}{4}\left(\frac{\partial\eta}{\partial\tau}\right)^{2}\\
\nonumber
&&-hx\hspace{0.1cm}\frac{\partial\eta}{\partial\tau}
-x+\frac{\eta^{4}}{\varepsilon^{4}}-\lambda\bigg].
\end{eqnarray}
See, for example, \cite{IVLEV1,IVLEV2}. The expression in the square brackets is the Lagrangian in terms of imaginary time.
 The coordinates $x(\tau)$ and $\eta(\tau)$ in Eq.~(\ref{27}) are solutions of the classical equations of motion 
\begin{equation} 
\label{28}
\frac{1}{2}\frac{\partial^{2}x}{\partial\tau^{2}}=-h\frac{\partial\eta}{\partial\tau}-1,\hspace{0.5cm}
\frac{1}{2}\frac{\partial^{2}\eta}{\partial\tau^{2}}=
-h\frac{\partial x}{\partial\tau}-\frac{4\eta^{3}}{\varepsilon^{4}}
\end{equation}
with the conditions 
\begin{equation} 
\label{29}
x(\tau_{0})=0,\hspace{0.3cm}\frac{\partial x}{\partial\tau}\bigg |_{\tau_{0}}=-2,\hspace{0.3cm}\eta(0)=0,
\hspace{0.3cm}\frac{\partial x}{\partial\tau}\bigg |_{0}=0.
\end{equation}
Since the trajectory terminates at the $\delta$ well, this leads to the first condition (\ref{29}). The second condition
(\ref{29}) results from Eq.~(\ref{4a}). The third condition (\ref{29}) corresponds to the physical exit point $y=0$. The
fourth condition (\ref{29}) is a property of an exit point where the electrons stops to move in the tunneling direction. 
Note, that the physical velocity perpendicular to the tunneling direction at the exit point from under the barrier is not
zero. In the dimensionless units used it is
\begin{equation} 
\label{29a}
\frac{\partial y}{\partial t}=-\frac{\partial\eta}{\partial\tau}\bigg|_{\tau_{0}}.
\end{equation}
A solution of Eqs.~(\ref{28}) corresponds to the total energy 
\begin{equation} 
\label{19}
\lambda=-\frac{1}{4}\left(\frac{\partial x}{\partial\tau}\right)^{2}
+\frac{1}{4}\left(\frac{\partial\eta}{\partial\tau}\right)^{2}-x+\frac{\eta^{4}}{\varepsilon^{4}}.
\end{equation}
The relation (\ref{19}) is the fifth condition to the equations (\ref{28}) which can be satisfied by a proper choice of
$\tau_{0}$. 
\subsection{Total action}
The trajectory terminates at the unphysical (complex) point $x=0$, $y=-i\eta(\tau_{0})$. One should connect this point with 
a physical one, for example, $x=0$, $y=0$. In other words, one should add to $A_{0}$ the action
\begin{equation} 
\label{20}
A_{1}=2B{\rm Im}\left[\sigma(0,-i\eta(i\tau_{0}))-\sigma(0,0)\right].
\end{equation}
to complete the tunneling exponent (\ref{14}). One can find the action (\ref{20}) by a direct solution of the 
Hamilton-Jacobi equation (\ref{4}). Since the condition 
(\ref{4a}) holds at all $y$, it follows from (\ref{4}) that
\begin{equation} 
\label{21}
\left(\frac{\partial \sigma(0,y)}{\partial y}\right)^{2}+\frac{y^{4}}{\varepsilon^{4}}=1+\lambda.
\end{equation}
The integration results in
\begin{equation} 
\label{35}
A_{1}=2B\varepsilon(1+\lambda)^{3/4}f\left[\frac{\eta(\tau_{0})}{\varepsilon(1+\lambda)^{1/4}}\right]
{\rm sgn}\left(\frac{\partial\eta}{\partial\tau}\bigg|_{\tau_{0}}\right),
\end{equation}
where
\begin{equation} 
\label{36}
f(z)=\int^{z}_{0}dv\sqrt{1-v^{4}}.
\end{equation}
\begin{figure}
\includegraphics[width=4.5cm]{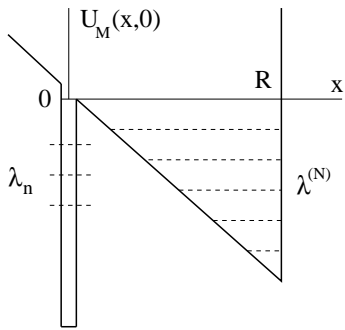}
\caption{\label{fig3}The modified potential with the infinite wall at $x=R$. The discrete energy levels $\lambda_{n}$ are 
associated with the $\delta$ well. The levels $\lambda^{(N)}$ are connected with the triangular well.}
\end{figure}
The sign term in Eq.~(\ref{35}) accounts for a correct sign of the square root just to match the solution at the point 
$x=0$, $y=-i\eta(\tau_{0})$. So the tunneling rate is given by
\begin{equation} 
\label{23}
w\sim\exp\left(-A_{0}-A_{1}\right),
\end{equation}
where the parts of the total action are determined by Eqs.~(\ref{27}) and (\ref{35}). 

It is not difficult to check that, if to formally drop down the part $y^{4}$ in the potential (\ref{2}), the trajectory 
formalism (\ref{23}) gives the same result (\ref{41}). For this purpose one has to write down the explicit expression for 
the classical trajectory
\begin{equation} 
\label{42}
\eta(\tau)= \frac{\sinh\left(2h\tau\right)}{h\sinh\left(2h\tau_{0}\right)}-\frac{\tau}{h},
\end{equation}
which follows from Eqs.~(\ref{28}), if to drop down the nonlinear part in $\eta$. The total underbarrier time is 
determined by the relation
\begin{equation} 
\label{43}
p\hspace{0.03cm}\sinh\left(2h\tau_{0}\right)=1.
\end{equation}
One should also put $f(z)=z$ in Eq.~(\ref{35}). At the terminal time $\tau_{0}$, as follows from Eqs.~(\ref{42}) and 
(\ref{43}),
\begin{equation} 
\label{44}
\eta(\tau_{0})=\frac{1}{h}-\frac{1}{2h^{2}}\ln\frac{1+\sqrt{1+p^{2}}}{p}.
\end{equation}
\section{NUMERICAL CALCULATIONS BY THE $X-Y$ NET}
\label{sec:masha}
For numerical studies of the problem we consider the Schr\"{o}dinger equation (\ref{26a}) at positive $x$ with the boundary
condition (\ref{4b}). Since in numerical calculations an infinite $x$ is impossible we modify the potential $U(x,y)$ 
(\ref{2}) adding the infinite potential wall at $x=R$. The wall is accounted for by the condition $\psi(R,y)=0$. In 
Fig.~\ref{fig3} the modified potential $U_{M}(x,0)$ is shown, where the electron is completely localized resulting in 
discrete eigenvalues of energy $\lambda$. In the $\delta$ well, due to the potential $y^{4}/\varepsilon^{4}$, there are
discrete levels $\lambda_{n}$. With no $y^{4}$ potential the set $\lambda_{n}$ becomes continuous. The levels associated 
with the triangular well are $\lambda^{(N)}$.

For the potential $U(x,y)$ (\ref{2}) an electron, initially localized at the $\delta$ well, can go to the infinite $x$ 
providing a decay of this metastable state during the typical time $t_{0}\sim 1/w_{T}$, where $w_{T}$ is the tunneling 
exponent (\ref{14}). 
\begin{figure}
\includegraphics[width=9cm]{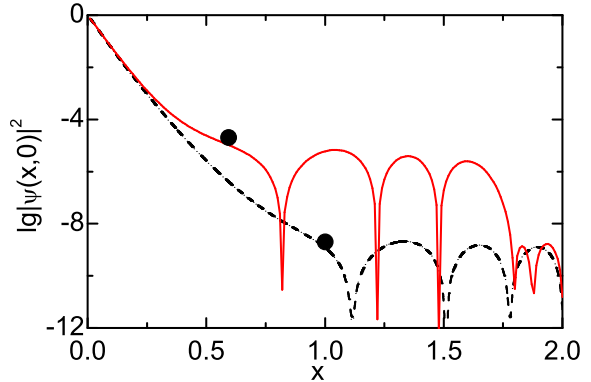}
\caption{\label{fig4}(Color online) Decimal logarithm of the electron density at $B=15$ calculated by the method of 
Sec.~\ref{sec:masha}. The dashed curve: Zero magnetic field, $h=0$. The solid curve: The finite magnetic field, $h=1.1$, 
and the eigenvalue at the wire $\lambda_{n}=-0.594$. The two dots mark the classical exit points calculated from 
Eqs.~(\ref{37}) and (\ref{38}).}
\end{figure}

For a state, initially localized at the $\delta$ well of the potential $U_{M}(x,y)$, the same type of time 
$t_{0}\sim 1/w_{T}$ is required to spread out to the triangular well when $\lambda_{n}$ is not equal to $\lambda^{(N)}$, 
that is we are away from Wigner resonances. The ratio of steady wave functions at the triangular well and at the $\delta$ 
well is exponentially small. If to avoid Wigner resonances, it is determined by the same exponent as the tunneling decay 
rate for the case of an infinite $R$. This justifies the use of $U_{M}(x,y)$.

When an energy level, occupied by the electron at the $\delta$ well, coincides with one at the triangular 
well the Wigner resonant tunneling occurs. In this situation the ratio (\ref{5}) of steady wave functions at the triangular
well and at the $\delta$ well is not exponentially small. Nevertheless, the time to fill out the initially empty triangular
well still remains exponentially long, $t_{0}\sim 1/\sqrt{w_{T}}$. 

We numerically calculated $\psi(x,y)$ for the potential $U_{M}(x,y)$ using the discrete two-dimensional net  in $\{x,y\}$ 
space. A step was chosen from a condition of a better convergence. The method used works well for hyperbolic system and
does not result in unphysical exponents. We started with the asymptotic $\psi\sim\exp\left(-By^{3}/3\varepsilon\right)$
at $y\rightarrow\infty$ and chosen eigenvalues of $\lambda$ from the condition of $\partial|\psi|/\partial y=0$ at $y=0$.
In the calculations we put $B=15$, $\varepsilon=1$, and $R=2$. The calculations are performed at the interval $0<x<2$ with 
the boundary condition (\ref{4b}) at $x=0$. A distance between energy levels in Fig.~\ref{fig3} can be estimated from the 
semiclassical approach \cite{LANDAU} as $(\lambda_{n+1}-\lambda_{n})\sim(\lambda^{(N+1)}-\lambda^{(N)})\sim 0.1-0.2$. This 
is confirmed by the numerical calculations. 

The results are figured out in Fig.~\ref{fig4}. 
\begin{figure}
\includegraphics[width=9cm]{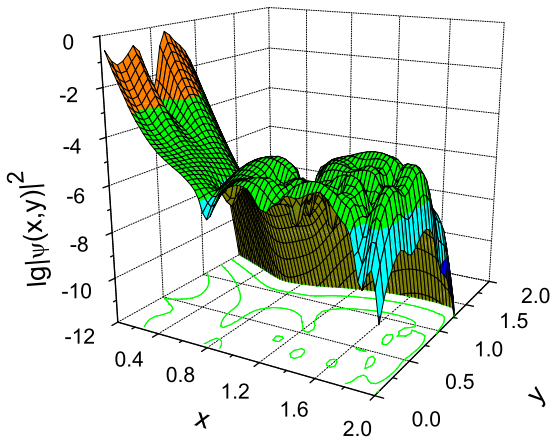}
\caption{\label{fig5}(Color online) The three-dimensional plot of the decimal logarithm of the electron density for
$h=1.1$, $\lambda=-0.594$, and $B=15$.}
\end{figure}

At zero magnetic field, $h=0$, the variables are separated and $\psi(x,0)$ is a solution of the effective Schr\"{o}dinger 
equation (\ref{36b}) with the potential (\ref{36c}), where one has to add the wall at $x=2$ and formally put $k=h=0$ and 
$\lambda\simeq -1$. The wave function is shown in Fig.~\ref{fig4} by the dashed curve. Due to reflections from the wall at 
$x=2$ the wave function oscillates at the interval $1<x<2$. At the classical exit point $x=1$ the wave function, calculated
by Eq.~(\ref{37}), is marked by the dot on the dashed curve in Fig.~\ref{fig4} where 
$|\psi(1,0)|^{2}\simeq 2\times 10^{-9}$.

At the finite magnetic field, $h=1.1$, the numerically calculated wave function is plotted by the solid curve in 
Fig.~\ref{fig4}. The chosen eigenvalue $\lambda=-0.594$ is one of the discrete levels $\lambda_{n}$ in Fig.~\ref{fig3}. The
wave function is extended over the whole interval $0<x<2$ due to reflections from the potential $y^{4}$ and the wall at 
$x=2$ in the magnetic field. A classical exit point $\{x(0),0\}$ and $\psi(x(0),0)$ can be calculated on the basis of the 
trajectory method of Sec.~\ref{sec:traj}. The results hardly differ from the case of Sec.~\ref{sec:wkb} when the potential 
$y^{4}$ was dropped down. According to Eq.~(\ref{44}), $\eta(\tau_{0})\simeq 0.263$ which leads to the very small potential
$y^{4}\simeq 0.005$. It is obvious and also can be shown that an influence of this potential on the classical trajectory is
small. We emphasize that in our case the potential $y^{4}$ hardly influences the underbarrier motion only. After an exit 
from under the barrier the electron is strongly reflected by that potential. 
\begin{figure}
\includegraphics[width=8cm]{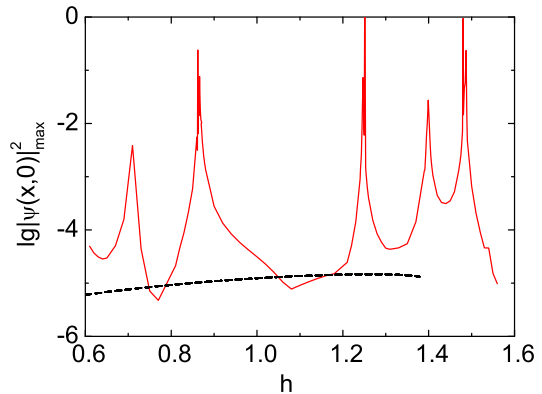}
\caption{\label{fig6}(Color online) Decimal logarithm of the maximal value of the wave function after an exit from under 
the barrier versus magnetic field $h$. The maxima relate to Wigner resonances, $\lambda_{n}=\lambda^{(N)}$. A level in the 
wire from the interval $0.55<\lambda_{n}<0.6$ is adjusted to the variable $h$. The dashed curve is $\lg w$ from 
Fig.~\ref{fig2} with $\lambda=-0.594$ for comparison. It corresponds to the electron density after the barrier when 
$R=\infty$ (no Wigner resonances).} 
\end{figure}

Neglecting the part $y^{4}$, one can easily perform trajectory calculations. This results in the exit point (\ref{38}),
$x_{0}=x(0)\simeq 0.594$, and the electron density $|\psi(x_{0},0)|^{2}\simeq 2.0\times 10^{-5}$ marked as the dot on the 
solid curve in Fig.~\ref{fig4}. We see that the analytical and numerical results are close to each other for both zero and 
the finite magnetic fields.

The three-dimensional plot of $|\psi(x,y)|^{2}$ is shown in Fig.~\ref{fig5}. Due to reflections from the potential $y^{4}$ 
and the wall at $x=2$, the motion covers the whole region apart the barrier. We also performed numerical calculations for 
the potential $y^{16}$ instead of $y^{4}$ just to be closer to a situation of an infinite restricting wall. Qualitatively,
the wave function looks the same filling out the whole $\{x,y\}$ space to the right of the barrier.

To get tunneling to the triangular well more adequate to tunneling to an infinite region, we should avoid Wigner 
resonances, as pointed out above. In Fig.~\ref{fig6} the maximal value of $|\psi(x,0)|^{2}_{max}$ after the exit point is 
plotted versus $h$. As one can see, the structure of Wigner resonances is very pronounced. A shape of the wave functions in
the triangular well close to Wigner peaks was checked. We found no deviation from eigenfunctions at the triangular well, 
when the $\delta$ well at $x=0$ was absent. This proves the Wigner nature of the peaks in Fig.~\ref{fig6}. We choose 
$h=1.1$ to get away of the Wigner resonances as clear from Fig.~\ref{fig6}.  

\section{NUMERICAL CALCULATIONS BY THE MATRIX FORMALISM}
\label{sec:vania}
In Sec.~\ref{sec:masha} we performed the numerical calculations at the interval $0<x<2$ using the boundary condition
(\ref{4b}) at $x=0$. In this section we demonstrate the numerical formalism, which does not explore the boundary condition
(\ref{4b}). The calculations are performed at the interval $-2<x<2$, where the potential $U_{M}(x,y)$ is
symmetrically continued from the region $0<x$ for negative $x$. The $\delta$ function is approximated by a narrow well of
the width of one computational step. The same method to model a $\delta$ well was used in Ref.~\cite{IVLEV8}. For positive 
$x$ in the semiclassical approach (a large $B$) the results for the symmetric potential chosen and for the potential
$U_{M}(x,y)$ should be close to each other. This is confirmed by the numerical calculations. 

In this section we use a matrix formalism. We start up with the Schr\"{o}dinger equation
\begin{eqnarray} 
\label{46}
&&-\left(\frac{1}{B}\frac{\partial}{\partial x}-ihy\right)^{2}\psi
-\frac{1}{B^{2}}\frac{\partial^{2}\psi}{\partial y^{2}}\\
\nonumber
&&+\left[\frac{y^{4}}{\varepsilon^{4}}-|x|-\frac{2}{B}\delta(x)\right]\psi=\lambda\psi
\end{eqnarray}
with the conditions $\psi(\pm 2,y)=0$. Eq.~(\ref{46}) differs from the form (\ref{26a}) by the gauge. The wave function can
be written as the expansion
\begin{equation} 
\label{47}
\psi(x,y)=\sum^{\infty}_{n=0}F_{n}(x)\varphi_{n}(y\sqrt{B}),
\end{equation}
where $\varphi_{n}(z)$ is a normalized eigenfunction of a harmonic oscillator,
\begin{equation} 
\label{48}
-\frac{\partial^{2}\varphi_{n}}{\partial z^{2}}+z^{2}\varphi_{n}=(1+2n)\varphi_{n}. 
\end{equation}
One can easily show that the functions $F_{n}(x)$ satisfy the equations
\begin{eqnarray} 
\label{49}
\nonumber
&&-\frac{1}{B^{2}}\frac{\partial^{2}F_{n}}{\partial x^{2}}
+\frac{2ih}{B^{3/2}}\sum^{\infty}_{m=0}\beta^{m}_{n}\frac{\partial F_{m}}{\partial x}\\
\nonumber
&&+\sum^{\infty}_{m=0}\left(\frac{h^{2}-1}{B}\gamma^{m}_{n} +\frac{1}{B^{2}\varepsilon^{4}}\alpha^{m}_{n}\right)F_{m}\\
&&+\left[\frac{1+2n}{B}-|x|-\frac{2}{B}\delta(x)-\lambda\right]F_{n}=0,
\end{eqnarray}
where the  matrices $\alpha^{m}_{n}$, $\beta^{m}_{n}$, and $\gamma^{m}_{n}$ are symmetric and 
\begin{equation} 
\label{50}
\alpha^{m}_{n}=\int^{\infty}_{-\infty}dz\varphi_{m}(z)z^{4}\varphi_{n}(z).
\end{equation}
The matrix $\beta^{m}_{n}$ ($\gamma^{m}_{n}$) is determined by the relation analogous to (\ref{50}) but with the 
substitution $z^{4}\rightarrow z$ ($z^{4}\rightarrow z^{2}$). The non-zero elements are
\begin{equation} 
\nonumber
\alpha^{n-4}_{n}=\frac{\sqrt{n(n-1)(n-2)(n-3)}}{4},
\end{equation}
\begin{equation} 
\nonumber
\alpha^{n}_{n}=\frac{3(2n^{2}+2n+1)}{2},\hspace{0.2cm}\alpha^{n-2}_{n}=\frac{(2n-1)\sqrt{n(n-1)}}{2}
\end{equation}
\begin{equation} 
\label{53}
\beta^{n-1}_{n}=\sqrt{\frac{n}{2}},\hspace{0.2cm}\gamma^{n-2}_{n}=\frac{\sqrt{n(n-1)}}{2},\hspace{0.2cm}
\gamma^{n}_{n}=\frac{1+2n}{2}
\end{equation}
and also $\alpha^{n+4}_{n}$, $\alpha^{n+2}_{n}$, $\beta^{n+1}_{n}$, and $\gamma^{n+2}_{n}$ obtained as symmetric 
combinations from (\ref{53}) with the shifts of $n$.
\begin{figure}
\includegraphics[width=7.2cm]{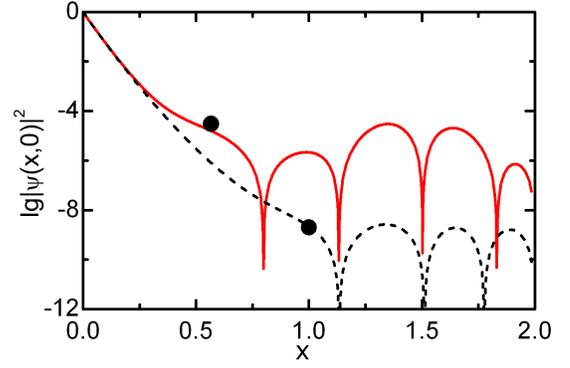}
\caption{\label{fig7}(Color online) Decimal logarithm of the electron density at $B=15$ calculated by the method of 
Sec.~\ref{sec:vania}. The dashed curve: Zero magnetic field, $h=0$. The solid curve: The finite magnetic field, $h=1.1$,
and the eigenvalue of the wire state $\lambda_{n}=-0.586$. The two dots mark the classical exit points 
$|\psi(1,0)|^{2}\simeq 2\times 10^{-9}$ for $h=0$ and $|\psi(0.586,0)|^{2}\simeq 2.3\times 10^{-5}$ for $h=1.1$ calculated 
from Eqs.~(\ref{37}) and (\ref{38}).}
\end{figure}

The above matrix formalism is used to numerically calculate the wave function (\ref{47}) by a solution of the eigenvalue 
problem (\ref{49}). The calculations were performed by means of the discretization of the interval $-2<x<2$ using 513
points. To check the scheme we made additional runs for the interval $-3<x<1$ with the equal number of discrete 
points. The both methods resulted in the same eigenvalues and eigenfunctions. In all the calculations we considered nine
terms in the expansion (\ref{47}) of the wave function.

The results are displayed in Fig.~\ref{fig7} showing the states generic with ones in Fig.~\ref{fig4}. The
difference in $\lambda_{n}$  for the both figures is 0.008 which is substantially less than 
$(\lambda_{n+1}-\lambda_{n})\sim 0.2$. 

Comparing Figs.~\ref{fig7} and \ref{fig4}, one can conclude that two different numerical methods lead to very similar 
results. The numerical data of the both methods are also in a good agreement with the trajectory results shown by the dots 
in Figs.~\ref{fig4} and \ref{fig7}.
\section{DISCUSSIONS}
\label{sec:disc}
A common point of view is that due to intrinsic underbarrier mechanisms a magnetic field always reduces a probability of 
tunneling decay of a metastable state. In other words, it never chances to get tunneling rate enhanced by a magnetic
field for many orders of magnitude. This point of view is supported by a general arguments that in the classical 
electrodynamics a motion ahead is prevented by cyclotron rotation and the same happens in quantum mechanics. The famous 
example is Landau states. The increased localization of a wave function in a magnetic field can also be explained by an
additional potential, proportional to squared velocity, created by the magnetic field. Any hand waving explanation leads to
a conclusion of tunneling reduction. We propose a mechanism which contrasts to that point of view.

Cyclotron enhancement of tunneling is clear for understanding. It occurs due to a rotation under the barrier of a velocity
vector of the electron. Under the cyclotron rotation it becomes more aligned with the tunneling direction, which increases 
the tunneling rate. A necessary condition for the phenomenon is a finite velocity (\ref{36e}) in the decayed state of the 
quantum wire. The velocity should be directed along the vector $\vec{\cal E}_{0}\times\vec H$, which is parallel to the 
wire. In the case of the opposite direction the magnetic field reduces the tunneling rate. For cyclotron enhancement of 
tunneling it is important that a sample, from where tunneling occurs, should be extended in the direction perpendicular to 
tunneling. The effect is absent in tunneling from a small quantum dot. 

The scenario of decay of the metastable state in the magnetic field consists of two parts. First, the electron moves under 
the barrier rotating its velocity. On this step, a role of reflections from the restricting walls is minor. Second, after
exit from under the barrier the electron goes to infinity performing multiple reflections from the restricting walls.

The semiclassical approach used is confirmed by two different numerical methods. A domain of parameters for the phenomenon 
can be roughly estimated from the condition that the Lorentz force, caused by the velocity in the wire, is of the order of 
the potential force ${\cal E}_{0}$. The parameter $h$ is 
\begin{equation} 
\label{45}
h=B\frac{\hbar\omega_{c}}{2u_{0}}\simeq 0.58\times 10^{-4}B\frac{H({\rm Tesla})}{u_{0}({\rm eV})}.
\end{equation}
Since we are interested in $h\sim 1$ and the semiclassical parameter $B$ is large, the Landau splitting $\hbar\omega_{c}$ 
is always less then $u_{0}$ which defines a scale of the energy levels in the quantum wire. 

To be specific, let us consider different regimes of tunneling.

{\it Strong effect on tunneling} ($h=2$). We take $B=40$ and $\lambda=-0.3$. Then at zero magnetic field the tunneling rate 
(\ref{41}) is $6.9\times 10^{-24}$. At $h=2.0$ the tunneling rate (\ref{41}) becomes $1.1\times 10^{-7}$. The enhancement 
occurs under the condition $u_{0}({\rm eV})\simeq 1.16\times 10^{-3}H({\rm Tesla})$. For reasonable values of the magnetic 
field, $H\simeq 10$~Tesla, the quantum wire or the thing film should be ``soft'' in the sense of a not large 
$u_{0}\sim 0.01$~eV. See Refs.~\cite{VDOV1,VDOV2,WIEL}. The exit point from under the barrier is estimated as 
$x_{0}\simeq 250\AA$ and the electric field is ${\cal E}_{0}\simeq 1.3\times 10^{3}$~eV/cm.

{\it Weak effect on tunneling} ($h=0.2$). We take $B=15.8$ and $\lambda=-0.3$. Then at zero magnetic field the tunneling rate 
(\ref{41}) is $7.1\times 10^{-10}$. At $h=0.2$ the tunneling rate (\ref{41}) becomes $1.0\times 10^{-7}$. The enhancement 
occurs under the condition $u_{0}({\rm eV})\simeq 1.16\times 10^{-2}H({\rm Tesla})$. For reasonable values of the magnetic 
field, $H\simeq 10$~Tesla, one can choose  $u_{0}\sim 0.1$~eV. The exit point from under the barrier is estimated as 
$x_{0}\simeq 75\AA$ and the electric field is ${\cal E}_{0}\simeq 1.0\times 10^{5}$~eV/cm.

One can conclude that an experimental observation of cyclotron enhancement of tunneling is possible. An experimental 
arrangement can be a quantum wire embedded into a two-dimensional electron system. The electric field is parallel to the 
two dimensional system and is perpendicular to the wire. The magnetic field is perpendicular to the wire and to the 
electric field. One can use a dielectric wall parallel to the tunneling direction just to get electrons moved away after 
tunneling along the wall due to multiple reflections from it in the magnetic field.

Another way of an experimental observation is to use a thin film, with a quantized motion inside, perpendicular to the 
electric field. 
\section{CONCLUSIONS}
A magnetic field can enhance a tunneling decay rate of a metastable state for many orders of magnitude. This happens due to
intrinsic underbarrier mechanisms of a cyclotron rotation of an electron velocity. It becomes more aligned with the 
tunneling direction resulting in the enhancement of the tunneling rate. The effect can be observed experimentally for 
tunneling from a quantum wire or a thing film across a barrier created by an applied electric field.
\acknowledgments
We thank S. V. Iordanski for valuable discussions. This work was partially supported by Russian Foundation for Basic 
Research and ``Financiadora de Estudos e Projetos - FINEP'' and Brazilian ``Minist\'erio da Ci\^encia e Tecnologia - MCT''.

\end{document}